\documentclass{iau}
\usepackage{graphicx}
\usepackage{epstopdf}

\title[Escaping cosmic rays and molecular clouds] 
{Interaction of escaping cosmic rays \\ with molecular clouds}

\author[S. Gabici]   
{Stefano Gabici$^1$}

\affiliation{$^1$APC, AstroParticule et Cosmologie, Universit\'e Paris Diderot, CNRS, CEA, Observatoire de Paris, Sorbonne Paris Cit\'e, France -- email: {\tt stefano.gabici@apc.univ-paris7.fr}}

\pubyear{2013}
\volume{xxx}  
\pagerange{119--126}
\jname{Title of your IAU Symposium}
\editors{A.C. Editor, B.D. Editor \& C.E. Editor, eds.}
\begin{document}

\maketitle

\begin{abstract}
The study of the gamma--ray radiation produced by cosmic rays that escape their accelerators is of paramount importance for (at least) two reasons: first, the detection of those gamma--ray photons can serve to identify the sources of cosmic rays and, second, the characteristics of that radiation give us constraints on the way in which cosmic rays propagate in the interstellar medium. 
This paper reviews the present status of the field.
\end{abstract}

\firstsection 
\section{Introduction}

The galactic disk is the sky's most prominent source of gamma rays in the GeV energy domain.  Discovered in the late sixties (\cite[Clark \etal\ 1968]{clark}), the galactic gamma--ray diffuse emission was soon interpreted as the decay of neutral pions ($\pi^0$) produced by cosmic rays (CRs) interacting with the interstellar gas (\cite[Stecker 1969]{stecker}). This confirmed a scenario first proposed by \cite[Hayakawa in 1952]{hayakawa} (for early reviews see \cite[Fazio 1967]{faziorew} and \cite[Ginzburg \& Syrovatskii 1964]{ginzburgsyrovatskii}). During the past decades the diffuse gamma--ray emission from the galactic disk has been observed with constantly increasing accuracy by several space instruments (\cite[Fichtel \etal\ 1975]{fichtel}, \cite[Mayer--Hasselwander \etal\ 1982]{cosBdiff}, \cite[Hunter \etal\ 1997]{egretdiff}, \cite[Ackermann \etal\ 2012a]{fermidiff}), and the spatial distribution of CRs in the Galaxy could be extracted from such observations (\cite[Bloemen 1989 and references therein]{bloemenrev}, \cite[Bertsch \etal\ 1993]{bertsch}, \cite[Strong \etal\ 2004]{stronggradient}, \cite[Ackermann \etal\ 2012]{fermidiff}). It turned out that the distribution of CRs in the Galaxy is quite uniform and that, as an order of magnitude, the intensity of CRs measured in the solar system is representative of the intensity anywhere else in the galactic disk. This roughly uniform background is often referred to as the {\it cosmic ray sea}.

The $\pi^0$--decay gamma--ray emissivity $q_{\gamma}^0(>100~{\rm MeV})$ of the local atomic gas has been measured by the {\it Fermi} LAT and is equal to $q_{\gamma}^0(>100~{\rm MeV})/4 \pi = 1.6 \times 10^{-26}$~ph/s/sr/H--atom (\cite[Abdo \etal\ 2009]{fermilocal}). Thus, an estimate of the intensity of the galactic diffuse emission from a specific direction in the sky can be obtained by integrating the gamma--ray emissivity of the gas $q_{\gamma}(>100~{\rm MeV}, l) \approx q_{\gamma}^0(>100~{\rm MeV})$ along the line of sight $l$. This straightforward procedure gives (\cite[e.g. Aharonian 2004]{felixbook}): 
\begin{equation}
\label{eq:diffuse}
J_{\gamma}(>100~{\rm MeV}) = \int {\rm d} l ~ \frac{q_{\gamma}(>100~{\rm MeV}, l)}{4 \pi} ~ n_{gas} ~\approx~ 1.6 \times 10^{-4} \left( \frac{N_H}{10^{22}~{\rm cm}^{-2}} \right) {\rm ph/s/cm^2/sr}
\end{equation}
where $N_H$ is a typical gas column density in the galactic disk.

In 1973 \cite[Black \& Fazio]{blackfazio} suggested that the gamma--ray emission from individual massive molecular clouds (MCs) might be visible above the diffuse galactic emission. The gamma--ray flux from a MC of mass $M_{cl}$, distance $d$ and embedded in the CR sea is:
\begin{equation}
\label{eq:cloud}
F_{\gamma}(>100~{\rm MeV})~ \approx~ \frac{q_{\gamma}^0(>100~{\rm MeV})~(M_{cl}/m_p)}{4 \pi ~d^2} ~\approx~ 2 \times 10^{-7} \left( \frac{M_{5}}{d_{kpc}^2} \right) ~{\rm ph/cm^2/s}
\end{equation}
where $M_{5}$ is the mass of the cloud in units of $10^5 M_{\odot}$, $d_{kpc}$ its distance in kiloparsecs, and $m_p$ the proton mass. The detectability of massive MCs above the diffuse galactic emission follows from Equations (\ref{eq:diffuse}) and (\ref{eq:cloud}) and from the fact that, for a typical cloud density of $n_{cl} \approx 1000~n_3$~cm$^{-3}$, the radius of the cloud is $R_{cl} \approx 10~ (M_{5}/n_{3})^{1/3}$~pc and its angular extension is $\Omega_{cl} \approx 10^{-4} (M_5/n_3)^{2/3} d_{kpc}^{-1/2}$~sr. 

\cite{blackfazio} proposed that the masses of the MCs of known distance could be derived from the strength of their gamma--ray flux , under the assumption that the intensity of CRs is known throughout the Galaxy (i.e. the assumption of a uniform CR sea). Such an approach was used extensively to calibrate the methods for the determination of the mass of molecular and atomic gas in the Galaxy and also to check the (rough) spatial homogeneity of the intensity of CRs in the Galaxy over large spatial scales (e.g. \cite[Caraveo \etal\ 1980]{caraveo}, \cite[Bloemen \etal\ 1984]{bloemencloud}, \cite[Hunter \etal\ 1994]{egretophiucus}, \cite[Digel \etal\ 1996]{egretcepheus}, \cite[1999]{egretorion}, \cite[Abdo \etal\ 2010a]{fermigould}, \cite[Ackermann \etal\ 2011]{fermiperseus}, \cite[2012b]{fermiorion}, \cite[2012c]{fermiclouds}, \cite[2012d]{fermicygnus}). However, the assumption of a uniform sea of CRs that permeates the whole Galaxy might be inappropriate in some circumstances, especially on small spatial scales (e.g. in the vicinity of CR sources). In these circumstances, the reasoning of Black \& Fazio can be reversed and, if an estimate of the mass of the cloud is available (see e.g. \cite[Hartquist 1983]{harquist} for caveats), one can use the gamma--ray observations of MCs to probe variations of the intensity of CRs in the Galaxy (\cite[Issa \& Wolfendale 1981]{issa}, \cite[Morfill \etal\ 1981]{morfill}, \cite[Aharonian 1991]{felixclouds}, \cite[2001]{felixclouds2}, \cite[Casanova \etal\ 2010]{sabrinabarometers}). In this context, MCs serve as {\it cosmic ray barometers}, because from their gamma--ray flux it is possible to infer the intensity (and thus pressure) of the CRs.

The fact that the study of MCs could help in solving the problem of the origin of CRs became evident when it was realized that an association between MCs and CR sources is indeed to be expected (\cite[Montmerle 1979]{SNOBs}). This is because CRs are believed to be accelerated at supernova remnant (SNR) shocks (e.g. \cite[Hillas 2005]{hillas}) and thus are likely to be produced in star forming environments (where core--collapse supernovae explode), which are in turn expected to host massive MCs. In a seminal paper, \cite{SNOBs} described a scenario in which CRs, after being accelerated at SNRs, could escape the acceleration site and diffuse to a nearby MC and produce there gamma rays via interactions with the gas. Due to the presence of these runaway CRs, the CR intensity is expected to be strongly enhanced in the vicinity of SNRs. In the same way, the gamma--ray emission from a MC illuminated by the runaway CRs will be much larger than the one derived in Equation~(\ref{eq:cloud}), which refers to a MC embedded in the CR sea.  It follows that the detection of gamma rays from MCs located in the vicinity of SNRs might constitute an hint for the fact that the nearby SNR is (or was, in the past) acting as a CR accelerator.

In this context, gamma--ray observations performed in the TeV domain are of great relevance, because: {\it i)} the expected gamma--ray emission for a MC illuminated by runaway CRs is expected to have a TeV flux which is within the reach of current Cherenkov instruments (Aharonian 1991, Aharonian \& Atoyan 1996, Gabici \etal\ 2009), and {\it ii)} in the TeV energy domain the contribution from the CR sea (which has a steep spectrum) to the gamma--ray emission is virtually negligible, and thus any detection of MCs has necessarily to be interpreted as an excess of CRs above the sea at the location of the MC (\cite[Aharonian 1991]{felixclouds}). A spectacular example of this fact can be found in \cite{hessridge}, where the detection of a diffuse emission of TeV gamma rays was reported from a very massive MC complex located in the galactic centre region. These observations revealed an excess above the CR sea of a factor of $\approx 4...10$ at TeV energies, and also an harder spectrum of CRs there (with slope $\approx 2.3$). The excess indicates that a source (or more sources) of CRs might be present in the region (remarkably, runaway CRs from only one SNR would suffice to explain the observed gamma--ray emission).

A phenomenological description of the propagation of the CRs after their escape from the acceleration site has been developed in a pioneering paper by \cite{aharonianatoyan}, who also discussed the expected radiative signatures (especially in gamma rays) due to the interactions of CRs in the ambient gas. Aharonian \& Atoyan considered an isotropic and spatially homogeneous diffusion coefficient for CRs and computed the expected gamma--ray emission from the MC, and stressed the fact that the properties of such emission (intensity, spectral shape, duration in time, etc.) strongly depend on the value of the diffusion coefficient. This opens the possibility to constrain, from gamma--ray observations, the diffusion coefficient of CRs in the vicinity of their sources. This fact has a tremendous importance for CR studies, given that the diffusion coefficient is a very poorly determined quantity (both from an observational and theoretical point of view).

The multiwavelength emission resulting from the interactions of runaway CRs in a MC has been computed, for the specific case in which a SNR accelerates the CRs, by \cite{me2007} and \cite{me2009}. These studies have then been applied to specific situations in order to obtain constraints on the particle diffusion coefficient in the vicinity of SNRs (\cite[Gabici \etal\ 2010]{me2010}, \cite[Nava \& Gabici 2013]{lara}). Moreover, the TeV diffuse emission resulting from the interactions of runaway CRs in the diffuse interstellar medium (i.e. in the absence of a massive MC) has been predicted and found to be within the reach of future ground based instruments such as the Cherenkov Telescope Array, which is thus expected to play a crucial role in proving (or falsifying) the SNR paradigm for the origin of galactic CRs  (\cite[Casanova \etal\ 2010]{sabrinaRXJ}, \cite[Acero \etal\ 2013]{fabio}). This paper is intended as a short review of these results.  A more extended discussion can be found in \cite{me2013}.

Though the effectiveness of the penetration of CRs into MCs still remains an open issue (\cite[Skilling \& Strong 1976]{skillingstrong}, \cite[Cesarsky \& V\"olk 1978]{cesarskyvolk}, \cite[Morfill 1982]{morfillpenetration}, \cite[Zweibel \& Shull 1982]{zweibel}, \cite[Everett \& Zweibel 2011]{everett}) I will assume in the following a full, unimpeded penetration of CRs in MCs. There is little doubt that TeV CRs can penetrate MCs (\cite[Gabici \etal\ 2007]{me2007b}) and the penetration of GeV particles seems to be supported by gamma--ray observations, from the early ones by \cite{lebrunpaul} to the ones by \cite{fermiorion}. For lower energies this remains an open issue, but this should not affect our considerations.

Here, I will not discuss the (yet not clear) way in which CRs escape their sources (see \cite[Gabici 2011]{escape} and references therein) nor the case in which the gamma--ray emission from the MC is the result of the interaction between the cloud itself and the SNR shock (numerous papers can be found in the literature, including: \cite[Blandford \& Cowie 1982]{cowie}, \cite[Aharonian \etal\ 1996]{adv}, \cite[Gaisser \etal\ 1998]{gaisserSNRs}, \cite[Bykov \etal\ 2000]{bykovIC443}, \cite[Fatuzzo \& Melia 2005]{fatuzzo}, \cite[Uchiyama \etal\ 2010]{yascrushed}, \cite[Malkov \etal\ 2011]{malkovbreak}, \cite[Inoue \etal\ 2012]{fukuiclouds}, \cite[Fang \& Zhang 2013]{fangzhang})

\section{Escape of cosmic rays from supernova remnants: isotropic diffusion}

As an illustrative example I consider here the case of the SNR W28, which is an aged remnant ($t_{age} \approx 4 \times 10^4$~yr) located in the vicinity of three massive MCs (of total mass $\approx 10^5 M_{\odot}$). Gamma--ray emission has been detected from the MCs in both the GeV and TeV domain (\cite[Aharonian \etal\ 2008]{hessw28}, \cite[Abdo \etal\ 2010b]{fermiw28}, \cite[Giuliani \etal\ 2010]{agilew28}). Since most of the gamma--ray emission clearly comes from {\it outside of the SNR shell}, it seems natural to interpret it as the result of the interactions in the MCs of CRs that escaped the SNR (\cite[Fujita \etal\ 2009]{fujita}, \cite[Gabici \etal\ 2010]{me2010}, \cite[Li \& Chen 2010]{lichen}, \cite[Ohira \etal\ 2011]{ohira}, \cite[Yan \etal\ 2012]{yan}). This supports the idea that W28 was, in the past, an accelerator of CRs.

I provide now a simple argument to show how one can attempt to constrain the diffusion coefficient in the vicinity of the SNR W28 by using the above mentioned gamma--ray observations, especially the ones performed by H.E.S.S. in the TeV domain. The time elapsed since CRs with a given energy $E$ escaped the SNR can be written as: $t_{diff} = t_{age} - t_{esc}$, where $t_{esc}(E)$ is the age of the SNR when CRs of energy $E$ were released. This is a time dependent quantity, since the highest energy CRs are believed to be released first, and CRs with lower and lower energy are gradually released at later times (Gabici 2011 and references therein). However, for CRs with energies above 1~TeV (the ones responsible for the very high energy gamma--ray emission) one can assume $t_{esc} << t_{age}$ (i.e. high energy CRs are released when the SNR is much younger than it is now) and thus $t_{diff} \sim t_{age}$. At time $t_{age}$ CRs have diffused over a distance $R_d \sim \sqrt{4 ~ D ~ t_{age}}$. Within the diffusion radius $R_d$ the spatial distribution of CRs, $f_{CR}$, is roughly constant, and proportional to $\eta E_{SN}/R_d^3$, where $E_{SN}$ is the supernova explosion energy and $\eta$ is the fraction of such energy converted into CRs. On the other hand, the observed gamma ray flux from each one of the three MCs detected in gamma rays is: $F_{\gamma} \propto f_{CR} M_{cl}/d^2$, where $M_{cl}$ is the mass of the MC and $d$ is the distance of the system. Note that in this expression $F_{\gamma}$ is calculated at a photon energy $E_{\gamma}$, while $f_{CR}$ is calculated at a CR energy $E_{CR} \sim 10 \times E_{\gamma}$, to account for the inelasticity of proton-proton interactions. By using the definitions of $f_{CR}$ and $R_d$ one can finally write the approximate equation, valid within a distance $R_d$ from the SNR:
$$
F_{\gamma} \propto \frac{\eta ~ E_{SN}}{(\chi ~ D_{gal} ~ t_{age})^{3/2}} \left( \frac{M_{cl}}{d^2} \right) .
$$
Estimates can be obtained for all the physical quantities in the equation except for the CR acceleration efficiency $\eta$ and the local diffusion coefficient $D$. By fitting the TeV data one can thus attempt to constrain, within the uncertainties given by the errors on the other measured quantities (namely, $E_{SN}$, $t_{age}$, $M_{cl}$, and $d$) and by the assumptions made (e.g. the CR injection spectrum is assumed to be $E^{-2}$, while the energy dependence of $D$ is assumed to scale as a power law of index $\delta = 0.5$), a combination of these two parameters (namely $\eta/D^{3/2}$). The fact that the MCs have to be located within a distance $R_d$ from the SNR can be verified a posteriori, and their exact location (unknown due to projection effects) can be tuned to match also the observed GeV emission. Given all the uncertainties above, our results have to be interpreted as a proof of concept of the fact that gamma ray observations of SNR/MC associations can serve as tools to estimate the CR diffusion coefficient. More detection of SNR/MC associations are needed in order to check whether the scenario described here applies to a whole class of objects and not only to a test-case as W28. Future observations from the Cherenkov Telescope Array will most likely solve this issue.

\begin{figure}
\begin{center}
 \includegraphics[width=0.32\textwidth]{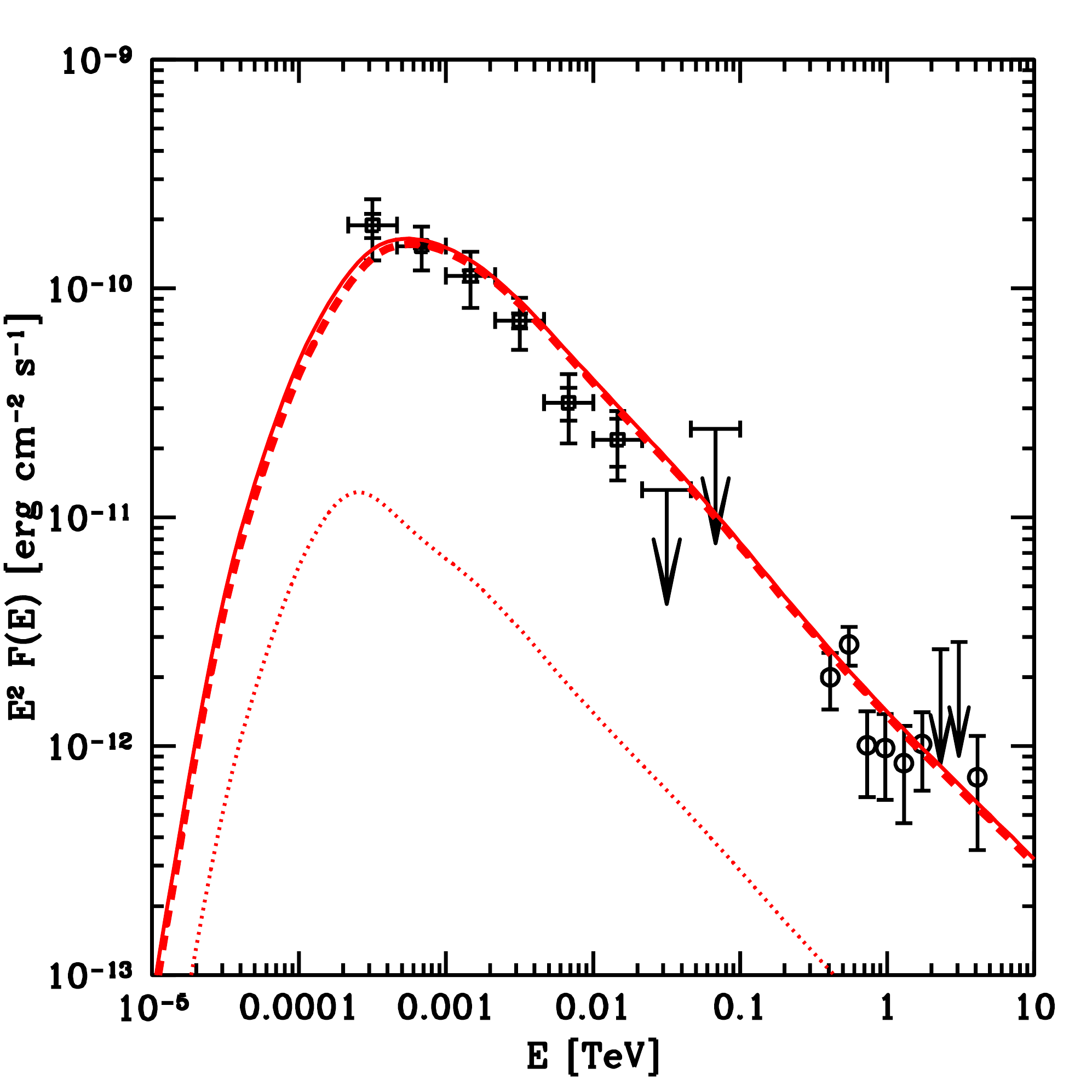} 
  \includegraphics[width=0.32\textwidth]{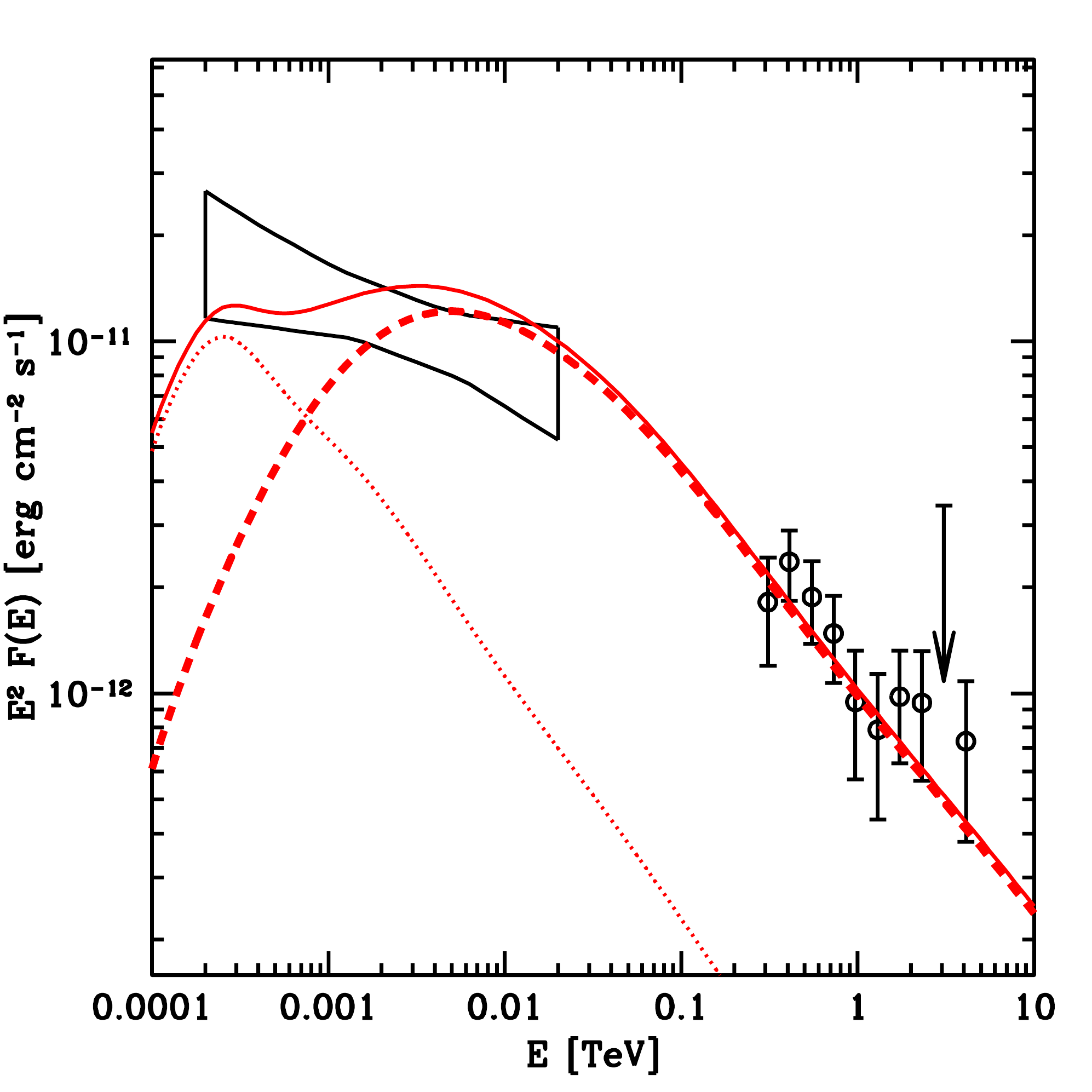} 
   \includegraphics[width=0.32\textwidth]{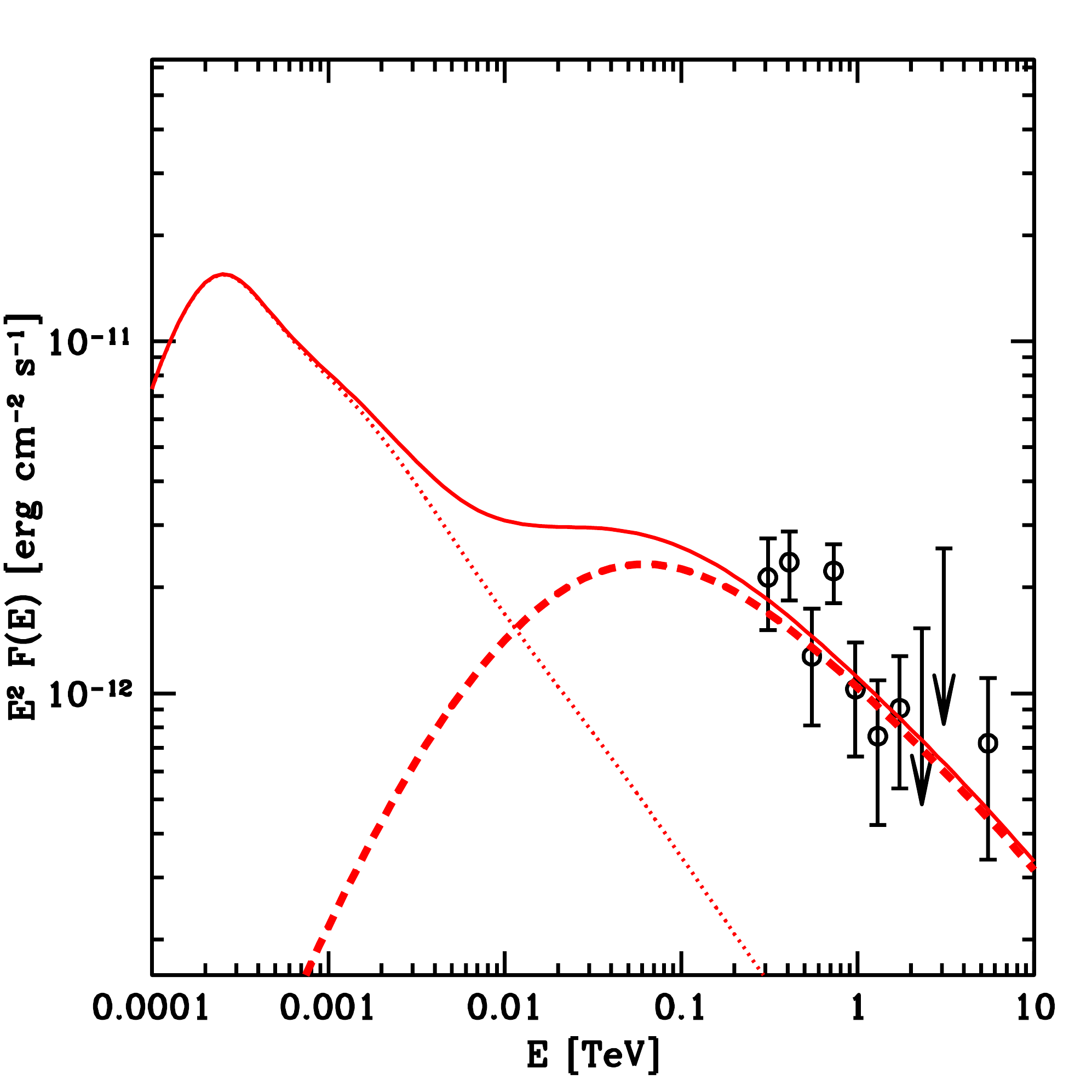} 
 \caption{Broad band fit to the gamma ray emission detected by {\it FERMI} and {\it HESS} from the sources HESS J1801-233, HESS J1800-240 A and B (left to right), that coincide with three massive MCs. Dashed lines represent the contribution to the gamma--ray emission from CRs that escaped W28, dotted lines show the contribution from the CR sea, and solid lines the total emission. Distances to the SNR centre are 12, 65, and 32 pc (left to right). {\it FERMI} and {\it HESS} data points are plotted in black. No GeV emission has been detected from HESS J1800-240 A.}
   \label{fig:1}
\end{center}
\end{figure}

Fig. \ref{fig:1}, from Gabici \etal\ 2010, shows a fit to the gamma--ray data for the three massive MCs in the W28 region. A simultaneous fit to all the three MCs is obtained by fixing a value for  $\eta/D^{3/2}$, which implies that the diffusion coefficient of particle with energy $3~{\rm TeV}$ (these are the particles that produce most of the emission observed by {\it HESS}) is:
\begin{equation}
D(3~{\rm TeV}) \approx 5 \times 10^{27} ~ \left(\frac{\eta}{0.1}\right)^{2/3} ~ {\rm cm^2/s} ~ .
\end{equation}
This value is significantly smaller (more than an order of magnitude) than the one normally adopted to describe the diffusion of $\sim$~TeV CRs in the galactic disk, which is $\approx 10^{29}~{\rm cm^2/s}$. This result remains valid (i.e. a suppression of the diffusion coefficient is indeed needed to fit data) even if a different value of the parameter $\delta$ is assumed, within the range 0.3...0.7 compatible with CR data. 

The reason of this discrepancy between the average CR diffusion coefficient in the Galaxy and the one found in the vicinity of a SNR needs to be explained. A possible way to interpret these observations is given in the next Section.

\section{Anisotropic diffusion of runaway cosmic rays}

Most of the studies aimed at predicting the gamma--ray emission from runaway CRs rely on the assumption of isotropic diffusion (see references in Sec.~2 and, e.g. \cite[Lee \etal\ 2008]{don2008}, \cite[Torres \etal\ 2008]{diegoIC443}, \cite[Rodr"guez Marrero \etal\ 2008]{diego2008}, \cite[Torres \etal\ 2010]{diego2010}, \cite[Ellison \& Bykov 2011]{don2011}, \cite[Li \& Chen 2012]{li2012}, \cite[Ellison \etal\ 22012]{don2012}, \cite[Telezhinsky \etal\ 2012]{igor2012}).
However, the validity of the assumption of isotropic diffusion of CRs, adopted in the previous section, needs to be discussed. In fact, if the intensity of the turbulent field $\delta B$ on scales resonant with the Larmor radius of particles is significantly smaller than the mean large scale field $B_0$ (i.e. if $\delta B/B_0 \ll 1$), then {\it cosmic ray diffusion becomes anisotropic}, with particles diffusing preferentially along the magnetic field lines (\cite[e.g. Casse \etal\ 2002 and references therein]{fabien}). In the limiting (but still reasonable) case in which the perpendicular diffusion coefficient can be set equal to zero, the transport of CRs across the mean field is mainly due to the wandering of magnetic field lines (\cite[Jokipii \& Parker 1969]{parker}). 

To give a qualitative idea of the role that anisotropic diffusion can play in the studies of the CRs that escaped SNRs, let us consider an idealized case in which the escaping particles diffuse along a magnetic flux tube characterized by a very long coherence length (i.e. the magnetic flux tube is preserved for a long distance). In this case, after a time $t$ particle will diffuse up to a distance $R_d \approx \sqrt{2\,D_\parallel \times t}$ along the tube (here $D_\parallel$ is the {\it parallel} diffusion coefficient of cosmic rays, not to be confused with the isotropic diffusion coefficient $D$ adopted in the previous Section), while their transverse distribution will be equal to the radius of the SNR shock at the time of their escape, $R_{sh}$, which is of the order of $\approx$~1--10 pc. Thus, the CR density in the flux tube will be proportional to $n_{CR} \propto (R_d R_{sh}^2)^{-1}$ instead of $\propto R_d^{-3}$ as in the isotropic case (see previous Section). It is easy to see that the estimates of the diffusion coefficient based on the two opposite assumptions of isotropic and one--dimensional diffusion will differ by a factor of $\approx (R_d/R_{sh})^{4/3}$, which can be much larger than an order of magnitude! Thus, it is of paramount importance to investigate how the interpretation of gamma--ray observations depends on the assumptions made concerning CR diffusion.

\begin{figure}
\begin{center}
 \includegraphics[width=0.8\textwidth]{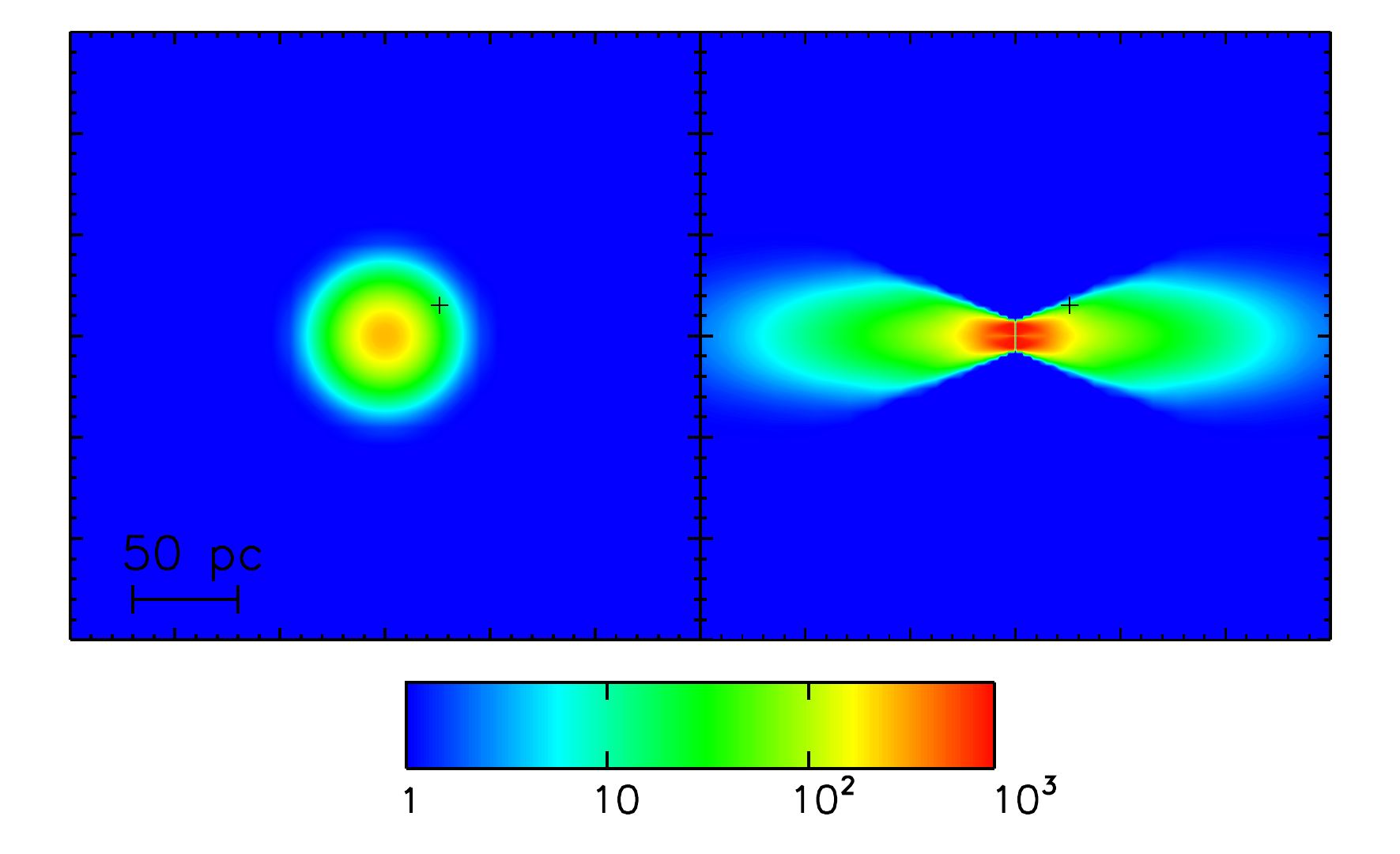} 
 \caption{Cosmic ray over-density above the galactic background around a supernova remnant (located at the centre of the panels). A particle energy of $E=1\,$TeV and a time $t=10\,$kyr after the explosion are considered. The left panel refers to an isotropic diffusion coefficient of cosmic rays equal to $D = 5 \times 10^{26} (E/10~{\rm GeV})^{0.5}\,$cm$^2$/s, while the right panel refers to an anisotropic diffusion scenario with $D_\parallel = 10^{28}(E/10~{\rm GeV})^{0.5}\,$cm$^2$/s. The black cross marks the a position at which the CR over-density is equal in the two panels. Figure from \cite{lara}.}
   \label{fig:2}
\end{center}
\end{figure}

As a first step, let us compare in Fig.~\ref{fig:2} the results that are obtained if an isotropic diffusion coefficient is assumed, with the ones obtained for the anisotropic diffusion model considered in Nava \& Gabici (2013). 
In both panels of Fig.~\ref{fig:2}, the SNR is located at the centre of the field and the color code refers to the excess of CRs with respect to the CR sea. 
Over-densities are plotted for a particle energy of $1\,$TeV and for a time $t = 10\,$kyr after the supernova explosion (see Nava \& Gabici 2013 for more details on the model).

The spatial distribution of CRs is strikingly different in the two scenarios: spherically symmetric in the left panel, and strongly elongated in the direction of the magnetic field flux tube in the right panel. A filamentary diffusion of CRs was also found in the numerical simulations by \cite{giacinti}.
The same parameters have been used to compute the over--densities in the two scenarios in Fig.~\ref{fig:2}, with the exception of the CR diffusion coefficient, which in the left panel has been assumed to be isotropic and equal to $D(1~{\rm TeV}) \approx 5 \times 10^{27}\,$cm$^2$/s, while in the right one is assumed to be strictly parallel (i.e. CRs diffuse only along field lines) and equal to $D_{\parallel} \approx 10^{29}\,$cm$^2$/s (a value similar to the average CR diffusion coefficient in the Galaxy). The choice of two significantly different values for the diffusion coefficients, with $D \ll D_\parallel$ has been made in order to obtain the same level of CR over--density in the vicinity of the SNR. As an example, the black cross in Fig.~\ref{fig:2} identifies a position, located $30\,$pc away from the centre of the explosion, where the CR over-density is identical in the two panels. 
To get comparable values for the CR over--density, a much smaller (isotropic) diffusion coefficient $D$ is needed in order to compensate for the larger solid angle over which CRs can propagate.
This fact must be taken into account when interpreting the gamma--ray observations of molecular clouds illuminated by CRs escaping from SNRs. For example, a fit to the gamma--ray data from the MCs in the W28 regions has been obtained by Nava \& Gabici (2013) by assuming a {\it large} diffusion coefficient of $D_{\parallel}(1~{\rm TeV}) \approx 10^{29}$~cm$^2$/s, much larger than the (isotropic) one adopted in the previous section (see Fig.~\ref{fig:1}). To conclude, the hint for a suppression of the diffusion coefficient in the vicinity of the SNR W28 obtained in the previous Section might depend on the assumption of isotropy of diffusion.
If an anisotropic diffusion is adopted, a much larger diffusion coefficient can be assumed to fit gamma--ray data.

\section{Conclusions and future perspectives}
\label{sec:conclusions}

I have shown how gamma--ray observations of MCs located close to SNRs can serve to support  the idea that SNRs are the sources of CRs. Information on the CR diffusion coefficient can also be extracted from such observations, though the conclusions of these studies strongly depend on the (still unknown) isotropic or anisotropic nature of diffusion. To date, only two SNRs show gamma--ray emission clearly coming from outside the SNR shell: W28 (see above) and W44 (\cite[Uchiyama \etal\ 2012]{yas2012}). So, further observations are needed in order to obtain solid constraints on the CR diffusion coefficient. Future facilities as the Cherenkov Telescope Array will play a key role in this direction.

Theoretical studies are also needed in order to understand the details of CR propagation close to their sources and interpret correctly the gamma--ray observations. The diffusion of CRs along the magnetic field lines is most likely a nonÐlinear process, where the CRs themselves generate the magnetic turbulence needed to confine them. This effect is expected to be stronger in the vicinity of CR sources, due to the enhanced intensity of CRs. Pioneering works on the non--linear propagation of runaway CRs can be found in \cite{skilling} and \cite{hartquistmorfill}. These studies have been revived by the recent results obtained from the gamma--ray observations of SNR/MC associations (see e.g. \cite[Ptuskin \etal\ 2008]{ptuskin}, \cite[Malkov \etal\ 2013]{malkov}), and promise to become one of the most important developments in this field.

\end{document}